\documentstyle[12pt,epsf]{article}
\begin{document}
\begin{titlepage}
\vskip 2cm
\begin{center}
{\Large\bf
CONSTRAINING FOUR GENERATION SM WITH $b\rightarrow
s\gamma$ AND  $b\rightarrow sg$ DECAYS}
\vskip 1cm  

{\large T. M. Aliev$^{a}$, D. A. Demir$^{b}$, N. K. Pak$^{a}$\\}
 
\vskip 0.5cm  
{\setlength{\baselineskip}{0.18in}
{\normalsize\it $^{a}$ Middle East Technical University, Department of 
Physics, 06531, Ankara, Turkey\\}
{\normalsize\it $^{b}$ Department of Physics, University of Pennsylvania, 
Philadelphia, PA 19104-6396\\}}
\end{center}
\vskip .5cm

\begin{abstract}
Using the experimental result on $b\rightarrow s\gamma$ and theoretical result on 
$b\rightarrow sg$, a four -generation SM is analysed to constrain the rephasing- 
invariant combinations of the CKM matrix and masses of the fourth generation quarks.
\end{abstract}
\end{titlepage}
\newpage

\section{Introduction}
Until now no experimental or theoretical proof for the absence of the extra 
generations of leptons and quarks has been given. The direct search for extra 
generations and tests of their indirect effects will be one of the major goals   
of the existing and next generation accelerators. For example, CDF [1] gives a 
lower bound of 139 GeV for the fourth generation quark masses assuming they are 
stable enough when leaving the detector. If the interaction with the detector is 
taken into account this lower bound falls by 30-40 GeV.  Thus, we can barely 
assume that the lower bound for the masses of the fourth generation quarks is 
somewhere around $M_{Z}$.

The fourth generation quarks and leptons have been considered in current literature 
due to various theoretical motivations. In [2] a four generation SM is proposed to 
solve  $\alpha_{s}$, $R_{b}$ and $R_{c}$ crises of SM at Z pole. As another 
solution to  these problems, in [3] a four  generation supersymmetric SM with 
$m_{t}\approx  M_{W}$ is proposed. In addition  to these, four generation SM has 
been used in  calculating the neutron Electric  Dipole Moment (EDM) (in the 
framework of the  Kobayashi-Maskawa Model) to break the  persisting $10^{-32} 
e-cm$ theoretical  boundary which is seven orders of magnitude less than the 
experimental upper bound.

Note that theoretically the smallness of the electroweak radiative corrections 
to the LEP1 observables, following Veltman's arguments, enables us to conclude   
that the masses of the fourth generation quarks and leptons must be almost
degenerate. This is the case because large mass splitting in the fourth
generation would induce unacceptably large loop corrections to LEP1 observables.

In this work we shall base our analysis on a four generation SM. We have two 
basic aims: 1) Determination of those rephasing invariant combinations of 
CKM matrix relevant to the calculation of the neutron EDM. 2) Foundation of some 
likelihood bounds for the fourth generation quark masses.

Denoting the 'top' and 'bottom' quarks of the fourth generation by $t'$ and $b'$
respectively, one can find three independent rephasing invariant combinations 
of the elements of the CKM matrix: $Im[V^{*}_{td}V_{tb}V^{*}_{cb}V_{cd}]$, 
$Im[V^{*}_{td}V_{tb}V^{*}_{t'b}V_{t'd}]$ and 
$Im[V^{*}_{ts}V_{tb}V^{*}_{t'b}V_{t's}]$. Other combinations can be expressed in 
terms of these and the moduli of the elements of the CKM matrix. As it was shown 
already in [4], strange quark dominates in the neutron EDM, and thus a calculation 
of $d_{s}$ yields at least an order of magnitude prediction for the neutron EDM.
In this sense our analysis is focused on the calculation of 
$Im[V^{*}_{ts}V_{tb}V^{*}_{t'b}V_{t's}]$.

For purposes mentioned above, in our analysis we shall make use of the 
experimental and theoretical results on the branching ratios of the 
FCNC decays  $b\rightarrow s\gamma$ and $b\rightarrow sg$.

We will first outline the derivations of the basic expressions for the 
quantities of interest.

Last section is devoted to numerical analysis and discussions. 
\section{Derivations}
$b\rightarrow s\gamma$ and $b\rightarrow sg$, being FCNC decays, start occuring  at 
the loop level of perturbation theory. These decays involve consecutive 
$b\rightarrow u, c, t, t'$ and $ u, c, t, t'\rightarrow s $ transitions. Thus
decay amplitude has the form 
\begin{eqnarray}
A=\sum_{f=u, c, t, t'}\lambda_{f}R(m^{2}_{f}/M^{2}_{W})
\end{eqnarray}
where $\lambda_{f}=V^{*}_{fs}V_{fb}$ and the function $R$
represents the combination of the results of the loop integrations and 
the relevant operators responsible for $b\rightarrow s\gamma$ and $b\rightarrow 
sg$ decays. The unitarity of the 4x4 CKM matrix gives 
\begin{eqnarray}
\lambda_{u}+\lambda_{c}+\lambda_{t}+\lambda_{t'}=0
\end{eqnarray}
It is known that $\lambda_{u}$ is much smaller than others, so from (2) it 
follows that 
\begin{eqnarray}
\lambda_{c} \approx -(\lambda_{t}+\lambda_{t'})           
\end{eqnarray}
Neglecting $m_{u}^{2}$ and $m_{c}^{2}$ compared to $M_{W}^{2}$, combining R's 
of different flavors with the help of (3), and taking the relevant operator 
structure for the decay under concern, one obtains  the following expressions 
for the $b\rightarrow s\gamma$ and $b\rightarrow sg$ decay amplitudes: 
\begin{eqnarray}
A_{b\rightarrow s\gamma} =\frac{G_{F}}{\sqrt{2}}[\lambda_{t} C_{7} + 
\lambda_{t'} C'_{7}]O_{7}\\
A_{b\rightarrow sg} =\frac{G_{F}}{\sqrt{2}}[\lambda_{t} C_{8} + 
\lambda_{t'} C_{8}']O_{8}.
\end{eqnarray}
where $O_{7}$ and $O_{8}$ are magnetic and chromo- magnetic penguin operators 
\begin{eqnarray}
O_{7}=\frac{e}{4\pi^{2}}m_{b}\bar{s}\sigma.F\,P_{R}b\\
O_{8}=\frac{g_{s}}{4\pi^{2}}m_{b}\bar{s}\sigma.G\,P_{R}b
\end{eqnarray}
Here $P_{R} = (1+\gamma_{5})/2$ and $F_{\mu\nu}$ and $G_{\mu\nu}$ are 
photon and gluon field strength tensors, respectively. The Wilson coefficients 
$C_{7}, C'_{7}, C_{8}$ and $C'_{8}$ in (4) and (5) are evaluated at the mass
scale $\mu\approx m_{b}$ using renorm group equations with five active 
flavours [5,6].
\begin{eqnarray}
C_{7}(\mu)=-\frac{1}{2}\eta^{16/23}A(x_{t}) 
-\frac{4}{3}(\eta^{14/23}-\eta^{16/23})B(x_{t})+\sum_{i=1}^{8}h_{i}\eta^{a_{i}}
\end{eqnarray}
\begin{eqnarray}
C_{8}(\mu)&=&-\frac{1}{2}(B(x_{t})-1.725)\eta^{14/23} -0.9135\eta^{0.4086} + 
0.0873\eta^{-0.4230}\nonumber\\&&-0.0571\eta^{-0.8994}+0.0209\eta^{0.1456}
\end{eqnarray}
where $x_{t}=m^{2}_{t}/M^{2}_{W}$, 
$\eta=\frac{\alpha_{s}(M_{W})}{\alpha_{s}(m_{b})}$, and 
\begin{eqnarray}
h_{i}&=&(2.2996,\,-1.088,\,-3/7,\,-1/14,\,-0.6494,\,-0.038,\,-0.0186,\,-0.0057)\\
a_{i}&=&(14/23,\,16/23,\,6/23,\,-12/23,\,0.4086,\,-0.423,\,-0.8994,\,0.1456)
\end{eqnarray}
$A(x)$ and $B(x)$ in (8) and (9) are given by
\begin{eqnarray}
A(x)&=&\frac{x(x-1)(8x^2+5x-7)+6x^2(2-3x)ln(x)}{12(x-1)^4)}\\
B(x)&=&\frac{x(x-1)(x^2-5x-2)+6x^2ln(x)}{4(x-1)^4)}
\end{eqnarray}
$C'_{7}$ and $C'_{8}$ could, respectively,  be obtained from (8) and (9) by 
replacing $x_{t}$ by $x_{t'}=m^{2}_{t'}/M^{2}_{W}$.
 
In order to minimize the $b$ -quark mass dependence, which leads to 
uncertainities, it is preferable to define braching ratios 
$BR(b\rightarrow s\gamma)$ and  $BR(b\rightarrow sg)$ by $BR^{exp}(b\rightarrow 
ce\nu)/\Gamma^{theor}(b\rightarrow ce\nu)$ times the decay rates calculated 
from the amplitudes in (4) and (5) where
$\Gamma^{theor}(b\rightarrow ce\nu)$ is given by [6,7]
\begin{eqnarray}
\Gamma^{theor}(b\rightarrow ce\nu)=\frac{G^{2}_{F}\mid \lambda_{c}\mid^{2}}
{192\pi^{3}\mid V_{cs}\mid^{2}}m^{5}_{b}g(m_{c}/m_{b})\kappa(m_{c}/m_{b})
\end{eqnarray}
with the phase space factor 
\begin{eqnarray}
g(x)=1-8x^{2}+8x^{6}-x^{8}-24x^{4}ln(x).
\end{eqnarray}
The factor $\kappa(x)$ describes the leading order (LO) QCD corrections and 
given by 
\begin{eqnarray}
\kappa(x)=1-\frac{2\alpha_{s}}{3\pi}((\pi^{2}-31/4)(1-x)^2+3/2).
\end{eqnarray}
Finally, upto two -loop accuracy where $\alpha_{s}(\mu)$ is given by 
\begin{eqnarray}
\alpha_{s}(\mu)&=&\frac{4\pi}{\beta_{0}L}(1-\frac{\beta_{1}ln(L)}{\beta_{0}^2 L})
\end{eqnarray}
where $L=ln(\frac{\mu^2}{\Lambda_{QCD}^2})$, $\beta_{0} =23/3$ and $\beta_{1} 
=116/3$, for five active flavours. Then $BR(b\rightarrow s\gamma)$ and 
$BR(b\rightarrow sg)$ can be written as  \begin{eqnarray}
BR(b\rightarrow s\gamma)&=&P_{\gamma}(C'_{7}-C_{7})^{2}[(R+R_{7})^{2} + I^{2}]\\
BR(b\rightarrow sg)&=&P_{g}(C'_{8}-C_{8})^{2}[(R+R_{8})^{2} + I^{2}] 
\end{eqnarray}
where the various symbols introduced for convenience have the meaning
\begin{eqnarray}
P_{\gamma}&=&\frac{6\alpha_{e}\mid V_{cs}\mid^{2} 
BR^{exp}(b\rightarrow ce\nu)}{\pi g(m_{c}/m_{b})\kappa(m_{c}/m_{b})}\\
P_{g}&=&\frac{8\alpha_{s}\mid V_{cs}\mid^{2}
BR^{exp}(b\rightarrow ce\nu)}{\pi g(m_{c}/m_{b})\kappa(m_{c}/m_{b})}\\
R_{7}&=&\frac{C_{7}}{C'_{7}-C_{7}}\\
R_{8}&=&\frac{C_{8}}{C'_{8}-C_{8}}\\
R&=&Re[\frac{\lambda_{t'}}{\lambda_{t}+\lambda_{t'}}]\\
I&=&Im[\frac{\lambda_{t'}}{\lambda_{t}+\lambda_{t'}}]
\end{eqnarray}
It is clear that $BR(b\rightarrow s\gamma)$ and $BR(b\rightarrow sg)$ depend
on the combination $\frac{\lambda_{t'}}{\lambda_{t}+\lambda_{t'}}$.
Introducing a polar representatinon for $\lambda_{t}$ and $\lambda_{t'}$ as
\begin{eqnarray}
\lambda_{t}&=&se^{i\theta}\\
\lambda_{t'}&=&s'e^{i\theta'}
\end{eqnarray}
a straightforward calculation yields 
\begin{eqnarray}
\frac{s'}{s}&=&\frac{\sqrt{R^2+I^2}}{\sqrt{(1-R)^2+I^2}}\\
 Sin\beta &=& \frac{I}{\sqrt{(R^2+I^2)((1-R)^2+I^2)}}
\end{eqnarray}
where $\beta =\theta-\theta'$. 

A representation of $\lambda_{t}$ and $\lambda_{t'}$ as in (26) and (27) 
is independent of particular parametrisation of the CKM matrix [8].
$BR(b\rightarrow s\gamma)$ in (18) contains theoretical uncertainities 
coming from $V_{cb}$, $\alpha_{s}(\mu)$, $BR^{exp}(b\rightarrow ce\nu)$ 
and quark masses. To reduce these uncertainities one can construct the 
ratio
\begin{eqnarray}
r = \frac{BR(b\rightarrow s\gamma)}{BR(b\rightarrow sg)}.
\end{eqnarray}

First solving (18) and (30) simultaneously for $R$ and $I$ and then using (28) 
and (29) one can express $s'/s$ and $Sin\beta$ in terms of $r$ and the parameters 
defined in equatinons (20-23).

The ratio $r$ in (30) is useful in reducing the theoretical uncertainities. 
Furthermore, since (18) and (19) are similar equations in their dependence on 
the parameters with uncertainities (except for $\alpha_{e}\rightarrow\alpha_{s}$ 
interchage) their theoretical calculation in any model cannot produce too different 
error bounds. It is known already that theoretical and experimental results for 
$BR(b\rightarrow s\gamma)$ are close to each other assuming specific values for 
CKM elements [7]. Therefore, one can safely regard $r$ to be free of theoretical 
and experimental uncertainities and it can be indentified with the ratio of the 
central value of  $BR^{exp}(b\rightarrow s\gamma)$ to that of $BR(b\rightarrow 
sg)$. The latter is  unknown, except for the SM prediction of 
$BR^{SM}(b\rightarrow sg)\sim (2.3\pm 0.6)\times 10^{-3}$ [9].  Thus throughout 
the analysis $r$ will be regarded a free parameter. 

>From these observations one can also predict the uncertainities in $s'$ and $\beta$.
>From (28) one observes that the uncertainity in $s'$ is determined mainly by $s$, 
namely the uncertainities coming from $BR^{exp}(b\rightarrow s\gamma)$ and other 
parameters in (18) contribute negligeably. However, as is seen from (29) the 
uncertainity in $\beta$ is of the same order as that in $I$.  

\section{Numerical Analysis and Discussions}
Table 1 shows the values of the parameters used in the numerical analysis. As is 
seen from this table $BR^{exp}(b\rightarrow s\gamma)$ is uncertain by $\sim 30\%$
which is much larger than the errors in other quantities. However, as is shown in 
[7] the theoretical result for the branching ratio of the hadronic decay 
$BR(B\rightarrow X_{s}\gamma)$ is uncertain by a similar amount. Thus, as long as 
$BR(B\rightarrow X_{s}\gamma)$ is represented by $BR(b\rightarrow s\gamma)$ with 
the LO Wilson coefficients in the  framework of the spectator model, the main 
source of uncertainity comes from the experimental result. This causes, in  
particular, the quantities $I$ and $R$, defined in (24) and (25), be uncertain by 
a similar amount as $BR^{exp}(b\rightarrow s\gamma)$. As it was argued at the 
end of the last section, $r$ in (30) depends mainly on the  central values of the 
branching ratios. After analyzing $s'$ and $Sin\beta$ for the central  values of 
the quantities in Table 1, the percentage error in their values will be set to 
that of $s$ and $I$, respectively. 

>From the Wofenstein parametrisation of the CKM matrix, $\mid \lambda_{t}\mid\sim 
\lambda^{2}$, where $\lambda\approx 0.22$ is the Cabibbo angle. In [10]  
it is argued  that $Im[\lambda_{t}\lambda_{t'}^{*}] \sim \lambda^{5}$ and 
Im$[V^{*}_{td}V_{tb}V^{*}_{cb}V_{cd}]$, Im$[V^{*}_{td}V_{tb}V^{*}_{t'b}V_{t'd}]$ 
$\sim \lambda^{7}$, which makes $Im[\lambda_{t}\lambda_{t'}^{*}]$ more important 
than the other two. Thus, to make comparison explicit, in the final graphical 
results we shall present the $r$ dependence of the function $g(r)$ defined by 
\begin{eqnarray}
g(r)=Im[\lambda_{t}\lambda_{t'}^{*}]/\lambda^{5}.
\end{eqnarray}

As it will be seen soon, for a fixed value of $m_{t'}$, Sin$\beta$ is 
very sensitive to the variations in $r$; Sin$\beta$ is meaningful (real
and in [-1, 1] interval) only for a limited range of $r$ values. 

If the addition of a fourth generation is an extension of SM, 
the allowed range of $r$ must include the SM prediction of $\sim 0.1$. This, 
in particular, follows from the sequental character of the families in case 
of which one does not expect the occurence of new operators in the associated 
decay amplitudes as summarized by (4) and (5).  In this sense, one can discard 
some range of $m_{t'}$ values that drives the SM prediction outside the window of 
the allowed values of $r$.

\subsection{Light Top Case}

The most important non-oblique correction to Z decays arises in the $Zb\bar{b}$
vertex which depends quadratically on $m_{t}$. Therefore, naively one expects 
$R_{b}$ to move toward the SM prediction for a lower value of $m_{t}$. In this 
context, in [3] a supersymmetric SM with $m_{t}\sim M_{W}$ and  $m_{t'}\sim 
m_{t}^{CDF}$ is proposed. Here supersymmetry is necessary to satisfy the 
requirements of the CDF signal. 

In this section we shall analyse $g(r)$ and Sin$\beta$ for $m_{t}\sim M_{W}$ and  
$m_{t'}\sim m_{t}^{CDF}$. In particular, we shall set $m_{t}=85 GeV$ and 
vary $m_{t'}$ in the CDF range. This $m_{t}$ and $m_{t'}$ values are away 
from representing the CDF data. In particular, dominance of the 
decay mode $t'\rightarrow b W^{+}$, and width of $t$ -quark  must be explained in 
some appropriate model, such as supersymmetry.  We leave aside this problem and 
analyze the light top quark case assuming its results might shed light on the 
construction of some extension of SM.

In Table 2 we summarize ,as $m_{t'}$ wanders in the CDF range, the minimum 
($r_{min}$) and maximum ($r_{max}$) values of $r$ between which all quantities 
are meaningful. We observe from this table that as $m_{t'}$ increases from 158 GeV
to 194 GeV the allowed range of $r$ gets narrower. Thus, as $m_{t'}$ approaches 
to the upper CDF limit, $r$ is forced to take values closer to the SM prediction of 
$r\sim 0.1$. If $m_{t'}$ is further increased $r_{max}$ approaches SM prediction, 
in particular, $r\sim 0.1$ is just left outside the allowed interval of $r$ 
values for $m_{t'}\approx 1.5 TeV$.

Fig.1 shows the variation of Im[$Sin\beta$] with $r$ for $m_{t}=85\,$GeV and  
$m_{t'}=176$ GeV. Sin$\beta$ becomes imaginary for $r<0.027560$ and $r>0.16766$ 
leaving a window around $r\sim 0.1$ as the appropriate region. 

Fig. 2 shows the variation of $g(r)$ in (31) with $r$. It takes the value 3.5 at 
$r=0.027560$ and vanishes at $r=0.16766$. In the rather narrow interval 
$0.027560<r<0.030$ $g(r)$ falls rapidly from 3.5 to unity.
Starting from unity at $r\sim 0.030$, it descents gradually to zero at $r=0.16766$. 
For $r\sim 0.1$, $g(r) \sim 0.2$ making 
Im$[\lambda_{t}\lambda^{*}_{t'}] \sim \lambda^{6}$ which is less than the 
assumtion made in [10]. However, for $r<\sim 0.030$, 
Im$[\lambda_{t}\lambda^{*}_{t'}]$ is well  above $\lambda^{5}$, approaching
$\lambda^{4}$ at the lower end. 

The errors in the input parameters 
effect the value of Im$[\lambda_{t}\lambda^{*}_{t'}]$ for a particular value of 
$r$. The allowed range of $r$ , The allowed range of $r$ , however, is not 
sensitive to uncertainities in the parameters in Table 1, as it varies with 
the central value of $BR(b\rightarrow sg)$ in a particular model.
This analysis is performed for the central values of  the input parameters . As 
mentioned before one expects, Sin$\beta$ be uncertain  by $\sim 30\%$, so the 
numerical results for Im$[\lambda_{t}\lambda^{*}_{t'}]$ have an uncertainity of 
the same order. 

\subsection{Heavy Top Case}
In this section we shall analyse $g(r)$ and Sin$\beta$ for $m_{t}=176\,GeV$ and
$m_{t'}\geq M_{Z}$. 

In Table 3 we summarize ,as $m_{t'}$ increases from $M_{Z}$ to 300 GeV, the minimum
($r_{min}$) and maximum ($r_{max}$) values of $r$ between which all quantities
are meaningful. We observe from this table that allowed range for $r$ is narrower 
than that in the light top quark case. Infact, as $m_{t'}$ exceeds 200 GeV SM 
prediction for $r$ is just left outside the allowed range of $r$. As $m_{t'}$ 
increases further the window of the allowed values of $r$ forgets completely the 
the SM expectation about $r$. One can therefore fairly say that for heavy top 
quark consistent with CDF results, mass of the top quark of the fourth generation 
must be somewhere in between $M_{Z}$ and $\sim 200\, GeV$. 

Fig.3 shows the variation of Im[$Sin\beta$] with $r$ for $m_{t}=176\,$GeV and  
$m_{t'}=150$ GeV. Sin$\beta$ becomes imaginary for $r<0.02716$ and $r>0.11284$
which permits a very narrow interval in which $r$ may take a value. 

Fig. 4 shows the variation of $g(r)$ with $r$. It takes the value $\sim 1.2$  at 
$r=0.027116$, then falls first rapidly to $\sim 0.2$ aroud $r\sim0.030$, and  
gradually to zero at $r\sim0.11284$. As we observe from this figure, $g(r)$ is
smaller than that in the light top case. Next, one concludes that  
Im$[\lambda_{t}\lambda^{*}_{t'}]$ is less than $\lambda^{5}$ almost throughout 
the interval. Especially, around $r\sim 0.1$, Im$[\lambda_{t}\lambda^{*}_{t'}]$
is around $\sim 0.04$, making it $\sim \lambda^{7}$.

For higher values of $m_{t'}$ the SM prediction of $r\sim 0.1$ is far outside the 
allowed region. This calculation is performed for the central values of 
the parameters in Table 1. As mentioned before one expects, Sin$\beta$ be 
uncertain by $\sim 30\%$, so the numerical results for 
Im$[\lambda_{t}\lambda^{*}_{t'}]$ has an uncertainity of the same order.

\subsection{Comparison of Two Cases and Conclusions}

If one assumes that $r$ does not depart from its SM value, rather clear 
expectations follow. In particular, in both cases,  it is not possible to draw 
Im$[\lambda_{t}\lambda^{*}_{t'}]$ to $\sim \lambda^{5}$. As is seen from Fig.2,
for $r=0.1$, Im$[\lambda_{t}\lambda^{*}_{t'}]\sim \lambda^{6}$. In the heavy top
quark case it is even smaller; as Fig.4 suggests, for $r=0.1$, 
Im$[\lambda_{t}\lambda^{*}_{t'}]\sim \lambda^{7.4}$. 

In both cases, as $m_{t'}$ increases, allowed range for $r$ squeezed to the 
lower end. In particular, for the light top case $r=0.1$ point is thrown to 
"no solution" solution for $m_{t'}\sim 1.5 TeV$ which is much larger than the 
$m_{t'}\sim 200 GeV$ -point of the heavy top quark case. However, choosing large
values for $m_{t'}$ does not guarantee that $g(r)$ takes values of the order of 
unity. Except for $r$ values around $r_{min}$, for any $m_{t'}$, however large,
generally $g(r)$ is much smaller than  unity, forcing 
Im$[\lambda_{t}\lambda^{*}_{t'}]<< \lambda^{5}$. Thus, consistency of large $m_{t'}$ 
and Im$[\lambda_{t}\lambda^{*}_{t'}] \sim \lambda^{5}$ assumption of [10]
is not generally satisfied except for the special point of $r\approx r_{min}$. 
This, in particular, requires a very large value for $BR(b\rightarrow sg)$ which 
is unlikely to occur under four -generation SM diagrammatics of  the $b\rightarrow 
sg$ decay. One thus concludes that, including the $30\%$ uncertainity in $g(r)$, 
it is unlikely to have Im$[\lambda_{t}\lambda^{*}_{t'}]\sim \lambda^{5}$ at any 
$m_{t'}$ in both cases.

In the light top quark case the window of the allowed range of $r$ values is wide.
Thus, it permits more deviations from the SM expectation. However, the SM 
prediction of $r\sim 0.1$ is well included in the window. In this sense, since 
$r\sim 0.1$ is a subset of the allowed $r$ values one has some extension 
of SM. Despite these results, the light top quark case (in this four generation 
SM form) is not acceptable phenomenologically since it is not consistent with 
CDF signal.

In the heavy top case, the allowed window of $r$ values is narrower than that of
the light top case. Thus, deviations from the SM prediction is not large as 
in the case of light top case. Width of the window changes rapidly with $m_{t'}$. 
Moreover, since for $m_{t'}>\sim 200\,GeV$ $r\sim 0.1$ point is thrown to 
"no solution" region of $r$ values, one can boldly say that $m_{t'}>\sim 
200\,GeV$ is unlikely  to be acceptable. It is in this sense that one is able to 
propose some  upper bound for $m_{t'}$. This analysis gives results only on 
$m_{t'}$, so fourth  generation lepton masses  and $m_{b'}$ are not restricted at 
all. There is no  phenomenological difficulty  with this case as long as the 
"almost- degeneracy"  condition explained in the  introduction is satisfied. If 
such a fourth sequental  family of leptons and quarks do indeed exist LEP1.5 or 
LEP2 will be able to detect them. 
\newline
\newline 
One of us (D.A.D.) thanks Scientific and Technical Research Council of Turkey for 
financial support. 

\newpage
\begin{table}[htbp]
\begin{center}
\begin{tabular}{||c|c|c|c|c|c|c|c||}
Parameter & Range
\\ \hline
$\mu$(GeV)& 4.8
\\
$m_{t}^{CDF}(GeV)$ & 176$\pm$ 18
\\
$\mid V_{cb}\mid$ & 0.9743$\pm$ 0.0007
\\
$BR^{exp}(b\rightarrow ce\nu)$ & (10.4$\pm0.4)\%$
 \\
$BR^{exp}(b\rightarrow s\gamma)$ & (2.32$\pm0.67)\,10^{-4}$
 \\
$\Lambda_{QCD}$(GeV) & 0.195$\pm$ 0.005
\\
$M_{W}$(GeV) & 80.33
\\
$m_{c}/m_{b}$ & 0.30$\pm$0.02
\end{tabular}
\end{center}
\caption{\label{table:back0}
Values of the input parameters used in the numerical analysis}
\end{table}
\begin{table}[htbp]
\begin{center}
\begin{tabular}{||c|c|c|c|c|c|c|c||}
$m_{t'}(GeV)$ & $r_{min}$ & $r_{max}$
\\ \hline
158 & 0.027523 & 0.18000
\\
176 & 0.027560 & 0.16766 
\\
194 & 0.027590 & 0.15765
\end{tabular}
\end{center}
\caption{\label{table:back1}
Boundaries of the allowed range of $r$ values for $m_{t}=85 GeV$.}
\end{table}
\begin{table}[htbp]
\begin{center}
\begin{tabular}{||c|c|c|c|c|c|c|c||}
$m_{t'}(GeV)$ & $r_{min}$ & $r_{max}$
\\ \hline
91 & 0.0274400 & 0.15903
\\
150 & 0.0271600 & 0.11284
\\
200 & 0.0269943 & 0.09644  
\\
300 & 0.0269500 & 0.08080
\end{tabular}
\end{center}
\caption{\label{table:back2}
Boundaries of the allowed range of $r$ values for $m_{t}=176 GeV$.}
\end{table}
\end{document}